\documentclass[aps,pra,reprint,superscriptaddress,showkeys]{revtex4-1}

\usepackage{subcaption}
\usepackage[normalem]{ulem}
\usepackage{graphicx}
\usepackage{amsmath,amsthm,mathtools}

\begin{document}

\author{Jakub Szlachetka}
\thanks{These authors contributed equally to this work}

\affiliation{Institute of Physics, Faculty of Physics, Astronomy and Informatics \\ Nicolaus Copernicus University, Grudziadzka 5, 87--100 Torun, Poland}

\author{Sylwia Kolenderska  }

\email{Corresponding author: skol745@aucklanduni.ac.nz}

\thanks{These authors contributed equally to this work}

\affiliation{The Dodd-Walls Centre for Photonic and Quantum Technologies, Department of Physics, University of Auckland, Auckland 1010, New Zealand}

\author{Piotr Kolenderski }
\affiliation{Institute of Physics, Faculty of Physics, Astronomy and Informatics \\ Nicolaus Copernicus University, Grudziadzka 5, 87--100 Torun, Poland}

\title{On using classical light in Quantum Optical Coherence Tomography}




\begin{abstract}
Quantum Optical Coherence Tomography (Q-OCT) presents many advantages over its classical counterpart, Optical Coherence Tomography (OCT): it provides an increased axial resolution and is immune to even orders of dispersion. The core of Q-OCT is quantum interference of negatively correlated entangled photon pairs obtained in a Hong-Ou-Mandel  configuration. This two-photon interference can be observed in the time domain in the form of dips or in the Fourier domain by means of a joint spectrum. The latter approach proved to be practical in the sense that it alleviated the requirement posed on light in Q-OCT to exhibit strict negative correlations, since the negative correlations  can be easily extracted in the Fourier domain as the main diagonal of the joint spectrum. In this work, we investigate the use of this spectral approach in which quantum interference is obtained with classical low-intensity light pulses. We report theoretical calculations and their experimental validation and show that although such classical light is much easier to launch into an experimental system, it offers limited benefits as compared to Q-OCT based on entangled light. We analyse the differences in the characteristics of the joint spectrum obtained with entangled photons and with classical light and explain the origins of these differences.
\end{abstract}

\maketitle

Quantum Optical Coherence Tomography (Q-OCT) \cite{abouraddy2002quantum, kolenderska2020quantum} is a non-classical counterpart of OCT, which is immune to the image-degrading chromatic dispersion and at the same time provides an enhanced resolution \cite{okano2013dispersion}. In this method, the traditional light source is replaced with a source of frequency-entangled photon pairs, and the traditional Michelson interferometer by the Hong-Ou-Mandel (HOM) interferometer \cite{hong1987measurement, branczyk2017hong}. The detection consists of two photodiodes which measure the rate of coincidence of photon’s arrival at the two output ports of a beamsplitter.

Q-OCT techniques can be divided into two types. The time-domain version of Q-OCT \cite{abouraddy2002quantum} is based on translating the reference arm mirror and detecting a dip in the signal, also known as HOM dip, whenever the reference arm’s length approaches the object’s arm length in the interferometer. The width of the dip determines the axial resolution and - due to the nature of frequency-entanglement of the photons - is 2 times smaller than a Point Spread Function peak in traditional OCT for the same spectral bandwidth.
Also, thanks to the negative frequency correlations of the photons, the width of the HOM dip is not affected by even orders of chromatic dispersion, which is enough to ensure optimal axial resolution over the axial scan depth.

In the Fourier domain version of Q-OCT (Fd-Q-OCT)\cite{kolenderska2020quantum}, reference arm mirror remains fixed, and two spectrometers measure wavelength-dependent coincidence rate producing a two-dimensional signal called the joint spectrum. To obtain an A-scan, a diagonal of the joint spectrum is extracted and Fourier transformed. Because a diagonal corresponds to perfect negative correlations, the resulting A-scan is not affected by even orders of dispersion. Not only is the joint spectrum faster to acquire, but also it inherently provides enough information for the removal of artefacts \cite{maliszewski2021artefact} - additional peaks which do not correspond to the object's structure and which ultimately scramble the image even for the simplest of objects.

Both types of Q-OCT use entangled photon pairs coming from SPDC and measure the effects of their quantum interference in two complementary domains. However, the quantum interference that lies at the heart of Q-OCT can also be observed with coherent states, so very weak light pulses. Here, we consider an applicability of such light in Fd-Q-OCT-like imaging and test it against SPDC-based Q-OCT in terms of resolution and dispersion cancellation. We characterise experimentally an Fd-Q-OCT system based on weak light pulses in terms of basic imaging parameters: the axial resolution, the imaging range and signal roll-off. We also present experimental results and supporting theoretical calculations, which clearly indicate that due to its probabilistic behaviour coherent states are not suitable to produce enhanced resolution and dispersion cancellation.

\begin{figure}[t]
\centering
\includegraphics[width=\linewidth]{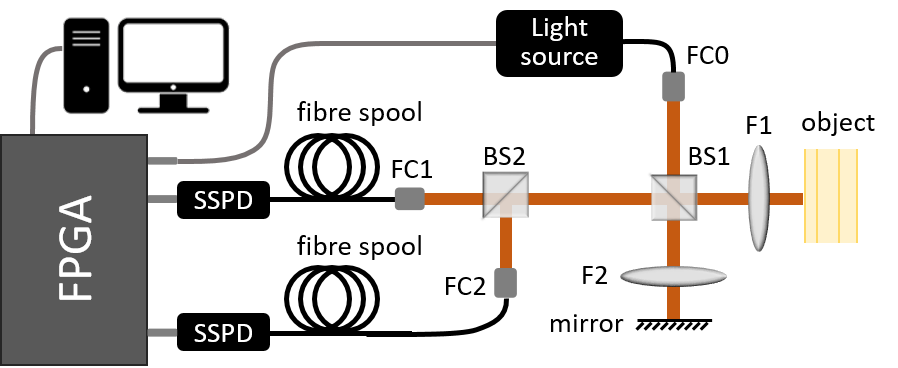}
\caption{A pulsed laser light attenuated to a level of single photons per pulse propagates in a Linnik-Michelson interferometer incorporating an object in the object arm and a mirror in the reference arm. The joint spectrum of the output of the interferometer is measured using a beamsplitter and two fibre spools with a time-resolving Superconducting Single-Photon Detector (SSPD) at their ends. Time reference is provided by a photodiode signal in the light source. The data is collected using an FPGA time-stamping electronics. F1, F2 - lenses, FC0-2 - fibre couplers, BS1-2 - 50:50 beamsplitters.}
\label{fig:setup}
\end{figure}

A schematics of a system used to test the imaging capabilities of Fd-Q-OCT based on coherent states is presented in Fig.~\ref{fig:setup}. Pulsed light with a central wavelength of 1550~nm and a total spectral bandwidth of 115~nm (MenloSystems T-Light) enters a Linnik-Michelson interferometer through a fibre collimator FC0 (f=11~mm). The interferometer consists of an object arm with a focusing lens F1 (f=50~mm) and a sample at its end and the reference arm with a focusing lens F2 (f=50~mm) and a mirror. Light reflected from an object and a reference mirror interferes at the same beamsplitter and goes on to the beamsplitter BS2. A photon arriving at BS2 can be either transmitted and propagate towards fibre collimator FC1 (f=11~mm) or it can be reflected towards fibre collimator FC2 (f=11~mm). Both FC1 and FC2  are connected to fibre spools (SMF28E, Fibrain) with the length of 5~km  and a group velocity dispersion, \( \beta_2 \), equal to 23~fs\(^2\)/mm. A fibre spool, through the phenomenon of dispersion, delays each wavelength by a different amount of time and therefore allows for spectrally-resolved measurements of the incident photons \cite{avenhaus2009fiber}.

The output ports of the fibre spools are monitored by Superconducting Single-Photon Detectors (SSPDs) (Scontel) whose  detection range is $350$ - $2300$~nm and a peak quantum efficiency - approximately 65\% at $1550$~nm \cite{Divochiy2018}. The SSPD outputs an electric pulse after each successful detection of a single photon and the FPGA (Field-programmable Gate Array) electronics measures the timestamps of the electric pulses. The timing jitter of the apparatus consisting of the SSPD and the time tagging unit is approximately $35$~ps. The light source was attenuated to a level of single photons per pulse by using a half-wave plate and a polarization beam-splitter at the input of the interferometer. The SSPDs were synchronized with the fast built-in photodiode in the light source to provide a joint spectrum of the light at the input of the fibre spools. Because the fibre spools' dispersion curve is not a linear function in wavenumber, a linearisation of the acquired spectra was performed \cite{szkulmowski2016spectrometer}.

\begin{figure}[t]
\centering
\includegraphics[width=\linewidth]{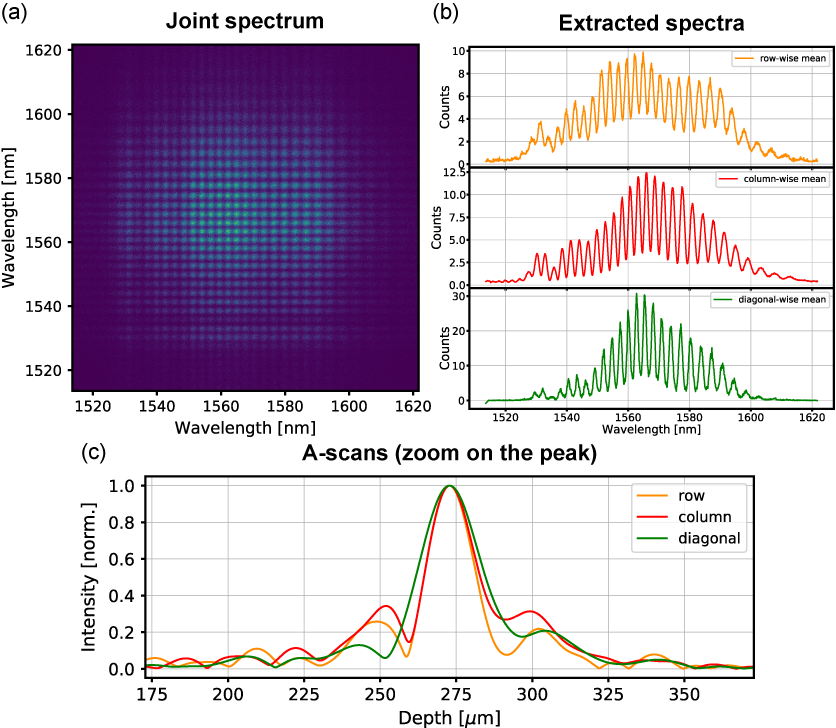}
\caption{\textbf{A mirror as an object placed at a position setting 273~\(\mu\)m optical path difference in the interferometer.} (a) A measured join spectrum; (b) Three types of spectra are extracted from the joint spectrum: a row-wise mean one (yellow) which corresponds to a spectrum which would be independently measured by one detection channel, column-wise mean one (red) that reflects the performance of the other detection channel, and a diagonal one (green), which is a mean of 20 central diagonals and corresponds to negatively correlated photons. (c) The comparison of widths of the peaks obtained after Fourier transforming the spectra shows no substantial difference in axial resolution.}
\label{fig:single_JS}
\end{figure}

We started by measuring a joint spectrum corresponding to the simplest object - a mirror placed at an optical path difference (OPD) in the interferometer equal to 273~\(\mu\)m. The result is depicted in Fig.~\ref{fig:single_JS}a. To test if the joint spectrum contains features that present an advantage over traditional OCT outputs, three different kinds of spectra were extracted (Fig.~\ref{fig:single_JS}b): a spectrum which is a mean over all rows of the joint spectrum, a spectrum being a mean of all the columns and finally and a spectrum being a mean of the central 20 diagonals of the joint spectrum. The first two spectra - row-wise and column-wise means - correspond to spectra of photons in an individual channel, so photons incident on fibre coupler FC1 or fibre coupler FC2, respectively. Although in theory, these spectra should have the same shape since they represent the same light source, in Fig.~\ref{fig:single_JS}b it can be seen that in practice the shapes slightly differ. This difference is related to the fact that it is experimentally impossible to achieve a perfect symmetry of the two detection channels. They will always slightly differ in terms of coupling efficiency, the induced dispersion, jitter of an SSPD etc., which will result in small dissimilarities between the spectra they produce.

All three spectra were first pre-processed to compensate the dispersion of the fibre spools and then Fourier transformed. The resultant A-scans are presented in Fig.~\ref{fig:single_JS}c, where an area around the peak is shown. The dispersion compensation needed to be performed separately for the three spectra due to the asymmetry of the detection channels. Consequently, initially the frequencies of the modulations on the pre-processed spectra can slightly differ, which - when Fourier transformed - gives a peak at slightly different position in the A-scan. However, this problem is partially solved when the X axis is properly recalculated to distances for all three cases as it is depicted in Fig.~\ref{fig:single_JS}c. The comparison in Fig.~\ref{fig:single_JS}c shows that the diagonal spectrum does not provide any resolution enhancement. In the presented example, the width of the peak corresponding to the diagonal spectrum is 23.3~\(\mu\)m and is bigger than the peaks corresponding to the row-wise mean spectrum, 17.6~\(\mu\)m, and column-wise mean spectrum, 18.7~\(\mu\)m. This resolution change is caused by the less favourable shape of the diagonal spectrum.

\begin{figure}[t]
\centering
\includegraphics[width=\linewidth]{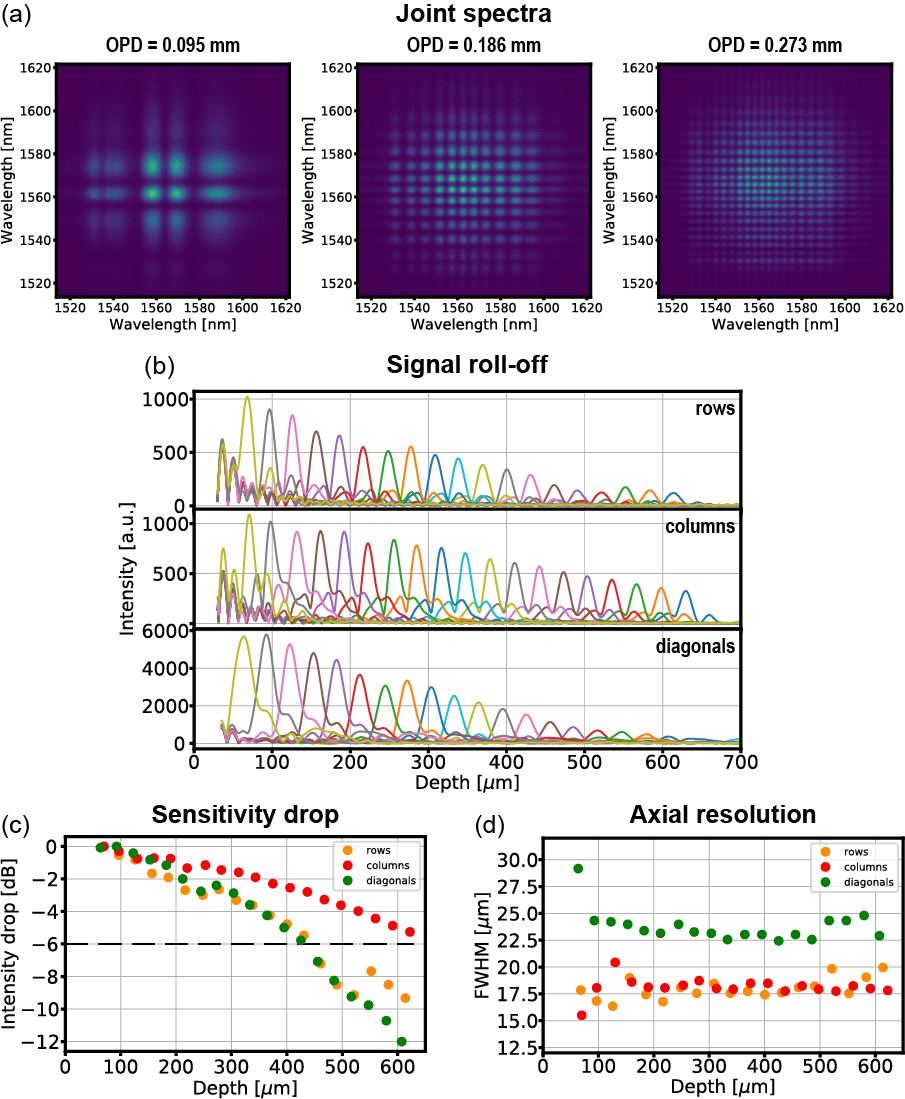}
\caption{\textbf{A mirror as an object placed at different positions.} (a) Joint spectra for three different OPDs in the interferometer; (b) signal roll-off corresponding to the three types of spectra that can be extracted from a joint spectrum; (c) the sensitivity drop for the three kinds of spectra shows a difference in performance of two detection channels; (d) the depth-dependent axial resolution remains at the same value for row- and column-wise means, but is slightly worse for the diagonal spectra due to their less favourable shape.}
\label{fig:im_range}
\end{figure}

In the next step, the mirror was axially translated and joint spectra were acquired at different OPDs (three of them are presented in Fig.~\ref{fig:im_range}a). Row-wise mean, column-wise mean and diagonal spectra were extracted from all the acquired joint spectra and dispersion compensated. Fourier transformation of these spectra gave three sets of A-scans with peaks whose position corresponds to an OPD in the interferometer. Each such set, which is shown in Fig.~\ref{fig:im_range}b, represents the signal roll-off, which visualises how the response of a spectrally-resolving interferometric system changes with the axial position of a reflective surface/a scatterer. For OCT systems, signal roll-off shows a depth-dependent decrease of the peak's height, which is caused by a drop in the fringes visibility on the spectrum due to a finite size of the sensor \cite{hu2007analytical}. Signal roll-off can be used to calculate a sensitivity drop - the height of the peak calculated as a \( 10 \log_{10}() \) at different depths (Fig.~\ref{fig:im_range}c). A distance after which the height of the peak decreases by 6 dB is known as an effective imaging range of an OCT system and determines the depth of an object at which high-quality imaging can be performed. The analysis of the sensitivity drop presented in Fig.~\ref{fig:im_range}c shows again a difference in the performance of the two detection channels: the 6-dB fall-off occurs at 0.63~mm for one (red dots) and at 0.44~mm for the other (orange dots). The sensitivity drop calculated for the diagonal spectra (green dots) is similar to the worse-performing channel with a 6-dB fall-off at 0.43~mm.

Finally, the axial resolution was calculated for all the three cases (Fig.~\ref{fig:im_range}d). As already preliminarly indicated in Fig.~\ref{fig:single_JS}c, it remains at a similar level for the two detection channels - approximately 17.5~\(\mu\)m throughout the effective imaging range for the row-wise mean spectra and 17.6~\(\mu\)m for the column-wise mean spectra. Due to a less favourable spectral shape, the FWHM of the peaks corresponding to the diagonal spectra is larger and gives a worse axial resolution, around 22.6~\(\mu\)m throughout the effective imaging range.

\begin{figure}[t]
\centering
\includegraphics[width=\linewidth]{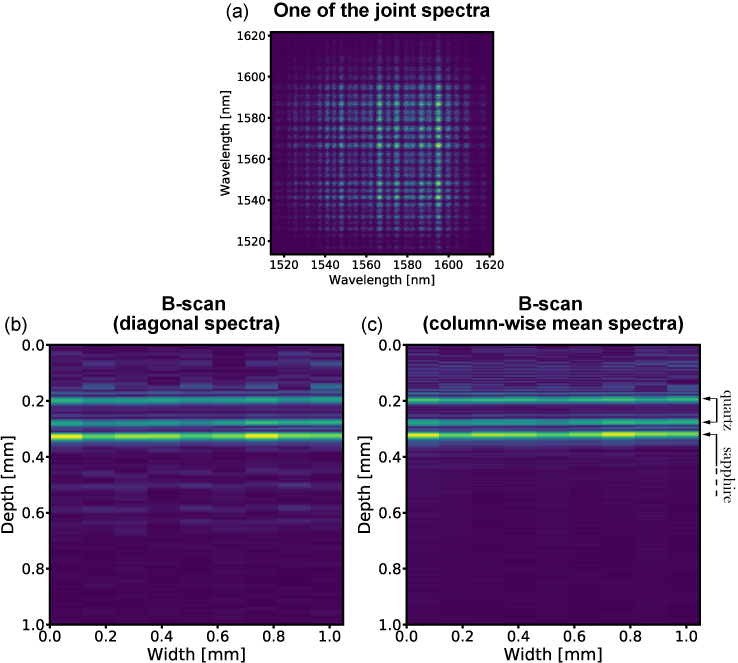}
\caption{ \textbf{A stack of glass as an object.} (a) One of the joint spectra acquired while laterally scanning the object; (b) a reconstructed B-scan for diagonal spectra and (c) a reconstructed B-scan for column-wise mean spectra show similar quality.}
\label{fig:glass}
\end{figure}

For a final confirmation of the similarity of the performance of traditional OCT with an OCT based on weak coherent states, a stack of glass - a 50\(\mu\)m thick quartz on top of a 300\(\mu\)m thick BK7 - was used as an object and a joint spectrum was acquired at 10 consecutive lateral positions of such an object. One of the acquired joint spectra (the fifth one) is presented in Fig.~\ref{fig:glass}a. Next, all the joint spectra were processed to generate two B-scans: one out of column-wise mean spectra (Fig.~\ref{fig:glass}c) and the other out of diagonal spectra (Fig.~\ref{fig:glass}b). Both images show a similar quality.

The coincidence rate measured in an experimental setup whose schematics is depicted in Fig.~\ref{fig:setup} can be modeled by the following probability function (see detailed calculations in supplementary materials):
\begin{gather}
     P_{cc}(\tau,\omega,\omega') =\nonumber \\
     = ( 1 - e^{- |\alpha(\omega) cos(\frac{\theta}{2})( f(\omega) cos^{2}(\frac{\theta}{2})e^{i 3 \phi_{t}}  -     e^{-i \omega \tau}  sin^{2}(\frac{\theta}{2}) e^{i \phi_{t}})|^{2} }) \nonumber\\ (1 - e^{- |\alpha(\omega') cos(\frac{\theta}{2}) sin^{2}(\frac{\theta}{2})( f(\omega')  e^{i(\phi_{t}-2\phi_{r})} + e^{-i \omega' \tau}  e^{-i(\phi_{t}+2\phi_{r})})|^{2} }),
     \label{mat:wzorek3}
\end{gather}

where  $\alpha(\omega) = \alpha u(\omega) $ with $\alpha$ being the mean number of photons and $u(\omega)$ - the light spectral amplitude, $\theta,\phi_{t},\phi_{r} $ correspond to the transformations performed by the beam-splitter \cite{campos1989quantum} and $\tau$ is the delay resulting from the difference in optical paths controlled by the change of the reference mirror position. $f(\omega)$ is the transfer function of the imaged object which for the remainder of this section we assume to be a mirror placed at a distance $z$ from a zero OPD point: $ f(\omega) = R e^{i z (\beta_{0} + \beta_{1} \omega)}$,  where \( R \) is the reflection coefficient, \( \beta_{0} \) - the wavenumber of light,  \( \beta_{1} \) - the inverse of the group velocity of light propagating in air. Next, eq. \eqref{mat:wzorek3} can be simplified by using $ \theta = \frac{\pi}{2}$, $\phi = 0$ and $\psi = \pi$: 
\begin{gather}
    P_{cc}(\tau=0,\omega,\omega') =\nonumber\\ ( 1 - e^{-\frac{\sqrt{2}}{4} |\alpha(\omega)|^2 |f(\omega)-1|^2} ) 
    ( 1 - e^{-\frac{\sqrt{2}}{4} |\alpha(\omega')|^2 |f(\omega')-1|^2} ).
    \label{mat:wzorek3-1}
\end{gather}
From the above equation, one can already deduce that the signal will not allow resolution-enhancement and dispersion cancellation. It is due to the term \( |f(\omega)-1|^2 \), which is responsible for the signal's modulatory character and which is also found in the equation describing a signal in traditional OCT \cite{kolenderska2021intensity}.

In the case of Fd-Q-OCT based on entangled photon pairs, the signal is expressed in the following form \cite{kolenderska2020quantum}:
\begin{gather}
    P_{qc} (\tau=0,\omega, \omega') = |\phi (\omega,\omega')|^{2} |f(\omega)  - f(\omega') |^{2} \\ \nonumber
    =|\phi (\omega,\omega')|^{2} \Big( |f(\omega) |^2 + |f(\omega')|^2 - 2 \mathrm{Re}\{ f(\omega) f^{*}(\omega')  \} \Big),
\end{gather}

\noindent where \( |\phi (\omega,\omega')|^{2} = \frac{1}{\sqrt{\pi \sigma \sigma'\sqrt{1-\rho^{2}}}}\exp{\left( \frac{\omega^{2}}{\sigma^{2}} + \frac{{\omega'}^{2}}{{\sigma'}^{2}} - \frac{2\omega \omega' \rho}{ \sigma \sigma'}\right)}\)  is the joint spectrum amplitude with \( \rho \) being the spectral correlation coefficient and \( \sigma \) and \( \sigma' \) - the spectral widths of the photons. \( f(\omega) f^{*}(\omega') \) in the last term in the bracket represents dispersion-cancelled and resolution-doubled structure of the object.

\begin{figure}[t]
     \centering
     \begin{subfigure}[b]{0.235\textwidth}
         \centering
         \includegraphics[width=1.1\textwidth]{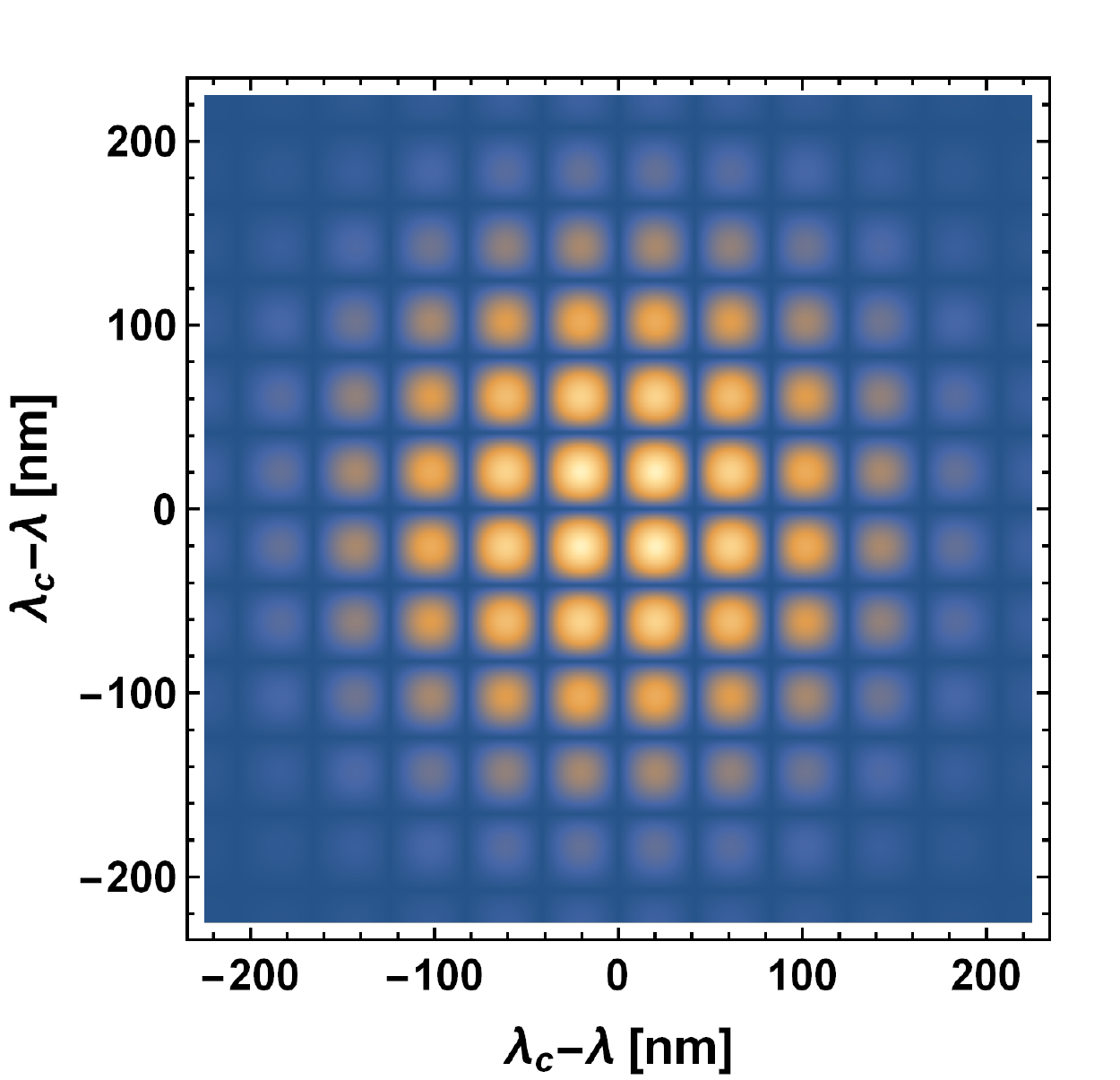}
         \caption{Coherent state}
         \label{fig:fig 5a}
     \end{subfigure}
     \hfill
     \begin{subfigure}[b]{0.235\textwidth}
         \centering
         \includegraphics[width=1.1\textwidth]{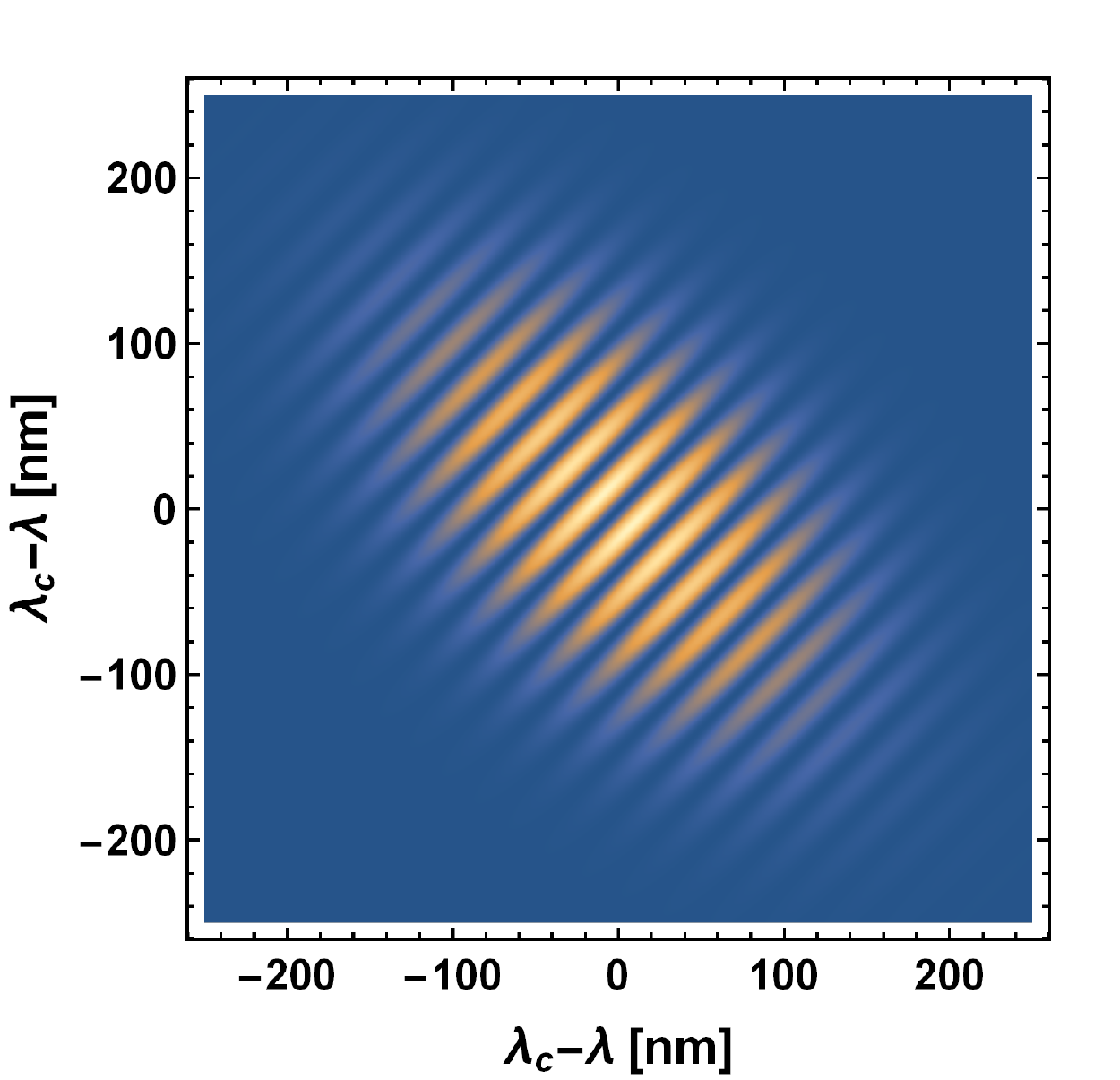}
         \caption{Biphoton state}
         \label{fig:fig 5b}
     \end{subfigure}
        \caption{Simulation of joint spectrum after light transmission through the experimental setup for an OPD of (a)  50 $\mu m$ in the Coherent state case and (b) 5 $\mu m$ in the Biphoton state case.}
        \label{fig:figure 5}
\end{figure}

Figure \ref{fig:figure 5} presents the joint spectrum produced  with attenuated light pulses ("Coherent sate" in fig. \ref{fig:fig 5b}) and with entangled photon pairs ("Biphoton sate" in fig. \ref{fig:fig 5b}). We assumed Gaussian spectra in both states and mirror as an object placed at $z$ = 5 $ \mu $m in Biphoton case and $z$ = 50 $ \mu $m in Coherent case  OPD. In the first case, the central wavelength $\lambda_{c}$ = 1550~nm, spectral widths of photons $\sigma$ , $\sigma'$ = 100~nm and the correlation parameter $\rho = -0.5$. 
In the second case, the optical parameters were the same except for \( \rho \), which was put as 0 to reflect the lack of correlation in the light. The parameter responsible for the transformation of the beam-splitter was $ \theta = \frac{\pi}{2}$ and corresponds to a transmission and reflection ratio of 50\%. Moreover, we assumed a phase $\phi = 0$, $\psi = \pi$ during the light passage through the beam-splitter. Additionally, the mean number of photons $\alpha = 0.1 $ is taken, which  indicates that  there is one photon per laser pulse.

The interference pattern in the joint spectrum for entangled photon pairs (Fig.~\ref{fig:fig 5b}) and the joint spectrum for attenuated light pulses (Fig.~\ref{fig:fig 5a}) looks differently. In the Biphoton state case, the interference pattern is superimposed on the joint spectrum at an angle with respect to the coordinate axes. However, in the Coherent state case, the interference pattern is perpendicular to the coordinate axes.

In summary, Fourier domain Q-OCT has undoubtedly opened up a new set of opportunities for Q-OCT imaging, but still remains very challenging experimentally as it requires entangled-photon-pairs sources which are not easy to set up. In this work, we considered an alternative source of light for the use in Fd-Q-OCT: a standard classical laser whose intensity is reduced to the level of single photons. The results of the conducted experiments show that such light used in an Fd-Q-OCT arrangement is not able to produce the advantages of standard Q-OCT: the resolution enhancement nor dispersion cancellation which are related to the spectral correlation within a photon pair. As seen in eq. \eqref{mat:wzorek3} and depicted in Fig.~ \ref{fig:figure 5} the joint spectrum is modulated by the interference of wave packets reflected form the reference  mirror and an object. This distinctive interference pattern does not originate from the spectral characteristics of the joint spectrum itself and in principle, could even be present in a correlation-free joint spectrum. Therefore, one could come to a conclusion that two photons from an attenuated laser pulse could be used for this purpose. However, there is also photon number correlation which needs to factored in. It is present in SPDC-generated pairs, but not in two independent coherent states, and results in the resolution enhancement and dispersion cancellation and leads to the distinctive pattern overlapping the joint spectrum. Our experimental and theoretical results confirm that the joint spectrum obtained with classical light - due to the lack of the correlation in the photon number - does not exhibit the advantages of the joint spectrum based on SPDC-generated pairs and as a result, merely reproduces the signal of traditional OCT signal in the horizontal, vertical and diagonal directions.

\medskip
\noindent\textbf{Funding Information.} SMK acknowledges The Dodd-Walls Centre for Photonic and Quantum Technologies (New Ideas Fund). PK and JS acknowledge financial support by National Science Centre, Poland, Sonata 12 grant no.~2016/23/D/ST2/02064 and National Laboratory of Atomic, Molecular and Optical Physics; 

\medskip

\noindent\textbf{Disclosures.} The authors declare no conflicts of interest.

\bibliography{On using classical light in Quantum Optical Coherence Tomography-arxiv}

\pagebreak
\widetext
\begin{center}
\textbf{\large Supplemental Materials: On using classical light in Quantum Optical Coherence Tomography}
\end{center}
\setcounter{equation}{0}
\setcounter{figure}{0}
\setcounter{table}{0}
\setcounter{page}{1}
\makeatletter
\renewcommand{\theequation}{S\arabic{equation}}
\renewcommand{\thefigure}{S\arabic{figure}}
\renewcommand{\bibnumfmt}[1]{[S#1]}
\renewcommand{\citenumfont}[1]{S#1}

\section{Derivation of the probability of a coincidence measurement}

\begin{figure}[t!]
\centering
\includegraphics[width=0.85\linewidth]{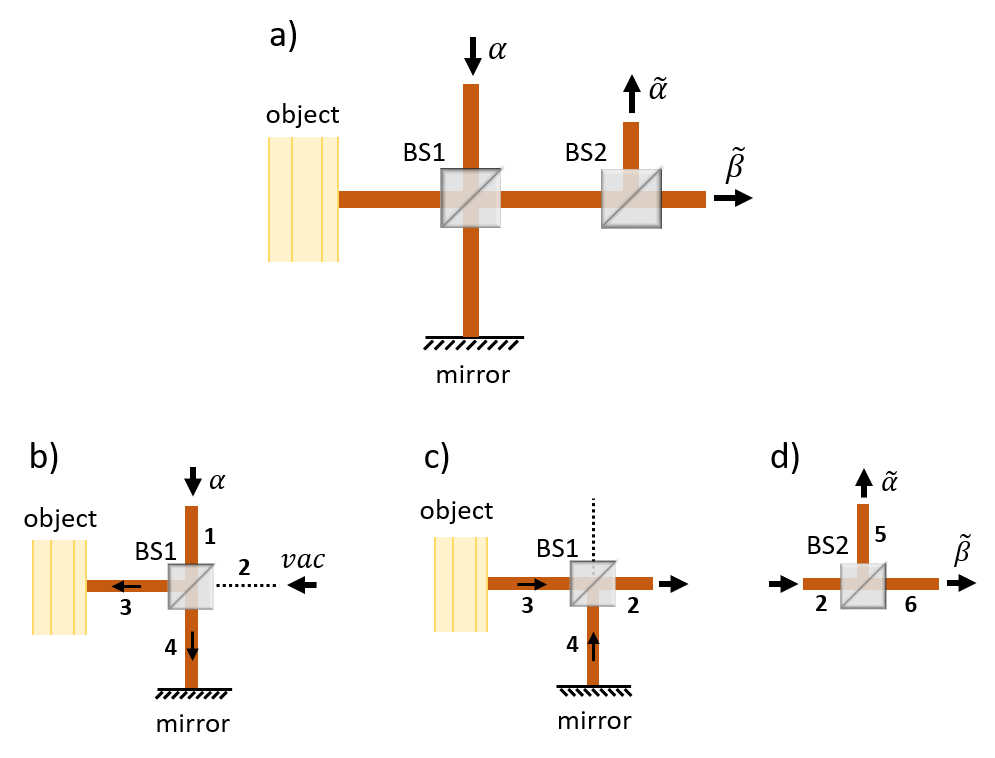}
\caption{\textbf{(a)} A scheme of the Michelson-Linnik interferometer setup. \textbf{(b)} A coherent state $\alpha$ propagates on path 1 to the first beam-splitter (BS 1). In path 2, there is no light (the lack of light is marked with vac). BS 1 splits the light into two paths, where the light propagates in the object (path 3) and reflects from the mirror (path 4). \textbf{(c)} The light reflected from the layers of the object and the mirror reach BS 1 which redirects it back to path 2. \textbf{(d)} The light on path 2 is incident on the second beam-splitter (BS 2), which splits it into two beams which are directed towards paths 5 and 6. As a result, states $\tilde{\alpha} $ and $ \tilde{\beta}$ are generated which are subject to measurement.}
\label{fig:setupz}
\end{figure}

This document aims at providing a formula expressing the result of a joint spectral measurement of weak light pulses in an Fd-Q-OCT setup. The light pulses are approximated by coherent states and are transformed by a a Michelson-Linnik interferometer with the object in the object arm and a mirror in the reference arm (Fig. \ref{fig:setupz}a).

We start by defining the transformation of a beam splitter in the form of a matrix equation [\textbf{Phys. Rev. A 40, 1371 (1989) }]:

\begin{equation}
   \left( \begin{array}{c}
{\hat{a}_{o}}^{+}(\omega)  \\
{\hat{b}_{o}}^{+}(\omega)  \\

\end{array} \right) = U(\theta, \phi_{t}, \phi_{r}) \left( \begin{array}{c}
{\hat{a}_{i}}^{+}(\omega)  \\
{\hat{b}_{i}}^{+}(\omega)  \\
\end{array} \right),
\label{mat:S1}
\end{equation}{}

\noindent where $\omega$ is the angular frequency,  ${\hat{a}_{i}}^{+}, {\hat{b}_{i}}^{+} $ are creative operators corresponding to the beam-splitter's input ports and  ${\hat{a}_{o}}^{+}, {\hat{b}_{o}}^{+} $ are creative operators corresponding to the beam-splitter's output ports. $U(\theta, \phi_{t}, \phi_{r})$ is the matrix representing the transformation performed by the beam-splitter and is written as: 

\begin{equation}
U(\theta, \psi, \phi) = \left( \begin{array}{cc}
cos(\frac{\theta}{2}) e^{i \phi_{t}} & sin(\frac{\theta}{2}) e^{i \phi_{r}}  \\
-sin(\frac{\theta}{2}) e^{-i \phi_{r}} & cos(\frac{\theta}{2}) e^{-i \phi_{t}}  \\

\end{array} \right) ,
\label{mat:S2}
\end{equation}

\noindent where $\theta$ is related to the transmittance and reflectance of the beam-splitter, $t$ and $r$, through the relationships $t = cos(\frac{\theta}{2})^2$ and $r = sin(\frac{\theta}{2})^2$. $\phi_{t}$ is a phase change due to the transmission by the beam-splitter and  $\phi_{r}$ is a phase change due to the reflection by the beam-splitter. To simplify the calculations, we introduce a beam-splitter transformation in the form of the following operator [\textbf{Opt. communications 62, 139–145 (1987)}]:

\begin{equation}
\hat{U}_{BS} =  e^{ - i (\phi_{t} - \phi_{r}) \hat{L}_{3}(\omega) } e^{ - i \theta \hat{L}_{2}(\omega) } e^{ - i (\phi_{t}+\phi_{r}) \hat{L}_{3}(\omega) },
\label{mat:Sbs}
\end{equation}
where $ \hat{L}_{3}(\omega) = \frac{1}{2} \left( \hat{a}_{i}^{+}(\omega) \hat{a}_{i}(\omega) - \hat{b}_{i}^{+}(\omega) \hat{b}_{i}(\omega)  \right) $  and $ \hat{L}_{2}(\omega) = \frac{1}{2 i} \left( \hat{a}_{i}^{+}(\omega) \hat{b}_{i}(\omega) - \hat{b}_{i}^{+}(\omega) \hat{a}_{i}(\omega)  \right) $ are analogues of angular momentum for the beam-splitter transformation  [\textbf{Phys. Rev. A 40, 1371 (1989) }] [\textbf{Phys. Rev. A 40, 1371 (1989) }].

The coherent state representing classical light used in the experiment is written as [\textbf{ Introductory quantum optics (Cambridge university press, 2005) }]:

\begin{equation}
     |\alpha \rangle = \hat{D}(\alpha) |vac \rangle = e^{-\frac{1}{2} \int d\omega |\alpha(\omega)|^{2}} e^{ \int d\omega \alpha(\omega) \hat{a}^{+}(\omega)} |vac \rangle, 
     \label{mat:S4}
\end{equation}{}

\noindent where $|vac\rangle$ is a state vector describing the vacuum, $\hat{D}(\alpha)$ is a displacement operator, $\alpha(\omega) = \alpha u(\omega)$ with $\alpha$ being the mean number of photons and $u(\omega)$ -  the light spectral amplitude.

At the input of the Michelson-Linnik interferometer, we propagate the laser light onto the first beam-splitter (BS1) port marked in the figure (Fig. \ref{fig:setupz}b) as path 1. In the second port of the beam-splitter, marked as path 2 in (Fig. \ref{fig:setupz}a), we assume vacuum. This situation can be written as:

\begin{gather}
    |\alpha\rangle_{1} |vac\rangle_{2} = e^{-\frac{1}{2} \int d\omega |\alpha(\omega)|^{2}} e^{ \int d\omega {\alpha(\omega)} {\hat{a}^{+}(\omega)}} |vac \rangle_{1} |vac \rangle_{2},
    \label{mat:S5}
\end{gather}{}

\noindent where the lower index next to the state vectors numbers the ports of interferometer.

BS1 transforms the state of \eqref{mat:S5} according to \eqref{mat:S1}, which can be expressed as an operator $\hat{U}_{BS}$ given in \eqref{mat:Sbs}. Because $\hat{U}_{BS}$ is unitary, $\hat{U}_{BS} {\hat{U}_{BS}}^{+} = 1 $, \eqref{mat:S5} can be rewritten as:

\begin{gather}
 e^{-\frac{1}{2} \int d\omega |\alpha(\omega)|^{2}} \hat{U}_{BS} e^{ \int d\omega {\alpha(\omega)} {\hat{a}^{+}(\omega)}}|vac\rangle_{1} |vac \rangle_{2} = \nonumber\\
    e^{-\frac{1}{2} \int d\omega |\alpha(\omega)|^{2}} \hat{U}_{BS} e^{ \int d\omega {\alpha(\omega)} {\hat{a}^{+}(\omega)}} {\hat{U}_{BS}}^{+} \hat{U}_{BS}|vac\rangle_{1} |vac\rangle_{2} = \nonumber\\
    e^{-\frac{1}{2} \int d\omega |\alpha(\omega)|^{2}} \Big(\hat{U}_{BS} e^{ \int d\omega {\alpha(\omega)} {\hat{a}^{+}(\omega)}} {\hat{U}_{BS}}^{+}\Big) \Big(\hat{U}_{BS}|vac\rangle_{1} |vac\rangle_{2}\Big)  .
    \label{mat:S6}
\end{gather}{}

The term in the first bracket of \eqref{mat:S6} will correspond to the transformation of the displacement operator  from the base of ports 1 and 2 to the base of ports 3 and 4, and the term in the second bracket will realise the transformation of the vacuum in ports 1 and 2 to the vacuum in ports 3  and  4. As a result, \eqref{mat:S6} can be rewritten as:

\begin{gather}
    e^{-\frac{1}{2} \int d\omega |\alpha(\omega)|^{2}} e^{{ \int d\omega {\alpha(\omega)} (cos(\frac{\theta}{2}) e^{i\phi_{t}}} {\hat{a}^{+}(\omega)} + sin(\frac{\theta}{2}) e^{-i \phi_{r}} {\hat{b}^{+}(\omega)}   ) } |vac\rangle_{3} |vac \rangle_{4}  = \nonumber \\
    |\alpha cos(\frac{\theta}{2}) e^{i \phi_{t}} \rangle_{3} |\alpha sin(\frac{\theta}{2}) e^{-i \phi_{r} } \rangle_{4} .
    \label{mat:S7}
\end{gather}{}

\noindent  The final form of the vector in \eqref{mat:S7} can be written in the presented form, if the sum of the square module of the coefficients at the creation operators is normalized to $|\alpha|^{2}$ and fulfills the definition of a coherent state expressed in \ref{mat:S4}:

\begin{gather}
    (\alpha cos(\frac{\theta}{2}) e^{i \phi_{t}} ) (\alpha cos(\frac{\theta}{2}) e^{i \phi_{t}})^{*}  + (\alpha sin(\frac{\theta}{2}) e^{i \phi_{r} })(\alpha sin(\frac{\theta}{2}) e^{i \phi_{r}})^{*} = \nonumber\\
    |\alpha|^{2} ({cos(\frac{\theta}{2})}^2 + {sin(\frac{\theta}{2})}^2) = |\alpha|^2 .
\end{gather}{}

The light on path 3 is reflected from the object with the transfer function $ f(\omega)$, and the light on the path 4 is reflected from the mirror whose position induces a time delay, $\tau$, between two arms of the interferometer and consequently, imparts a phase change $ e^{- i \omega \tau}$. The state of light after such transformations is:

\begin{equation}
     e^{-\frac{1}{2} \int d\omega |\alpha(\omega)|^{2}} e^{{ \int d\omega {\alpha(\omega)} (cos(\frac{\theta}{2}) e^{i \phi_{t}}} {\hat{a}^{+}(\omega)} f(\omega)} e^{{ \int d\omega {\alpha(\omega)} e^{- i \omega \tau} sin(\frac{\theta}{2}) e^{-i \phi_{r}}} {\hat{b}^{+}(\omega)}}|vac \rangle_{3} | vac\rangle_{4} .
     \label{mat:S9}
\end{equation}{}

After the reflections in the arms of the interferometer, the light from paths 3 and 4 is incident on BS~1 again (Fig. \ref{fig:setupz}c). Just as before, we use the operator given by the formula \eqref{mat:Sbs} and the properties of the unitary operator $\hat{U}_{BS} {\hat{U}_{BS}}^{+} = 1 $: 

\begin{gather}
e^{-\frac{1}{2} \int d\omega |\alpha(\omega)|^{2}}\hat{U}_{BS} e^{{ \int d\omega {\alpha(\omega)} (cos(\frac{\theta}{2}) e^{i\phi_{t}}} {\hat{a}^{+}(\omega)} f(\omega)} e^{{ \int d\omega {\alpha(\omega)} e^{- i \omega \tau} sin(\frac{\theta}{2}) e^{-i \phi_{r}}} {\hat{b}^{+}(\omega)}}|vac \rangle_{3} | vac\rangle_{4} = \nonumber\\
e^{-\frac{1}{2} \int d\omega |\alpha(\omega)|^{2}} \Big( \hat{U}_{BS} e^{{ \int d\omega {\alpha(\omega)} (cos(\frac{\theta}{2}) e^{i \phi_{t} }} {\hat{a}^{+}(\omega)} f(\omega)}{\hat{U}_{BS}}^{+} \Big) 
\Big( {\hat{U}_{BS}}  e^{{ \int d\omega {\alpha(\omega)} e^{- i \omega \tau} sin(\frac{\theta}{2}) e^{-i \phi_{r}}} {\hat{b}^{+}(\omega)}}{\hat{U}_{BS}}^{+}\Big) \times \nonumber\\
\times \Big(\hat{U}_{BS}|vac \rangle_{3} | vac\rangle_{4} \Big).
\end{gather}

Note that the expression  $ \hat{U}_{BS}|vac\rangle_{3}|vac\rangle_{4} =>|vac\rangle_{2}|vac\rangle_{1}$ transforms vacuum in ports 3  and 4 to the vacuum in ports 1 and 2. In the next step, the displacement operator is transformed from the base of ports 3 and 4 to the base of ports 1 and 2:

\begin{gather}
e^{-\frac{1}{2} \int d\omega |\alpha(\omega)|^{2}}\hat{U}_{BS} e^{{ \int d\omega {\alpha(\omega)} (cos(\frac{\theta}{2}) e^{i \phi_{t}}} {\hat{a}^{+}(\omega)} f(\omega)}{\hat{U}_{BS}}^{+} {\hat{U}_{BS}}  e^{{ \int d\omega {\alpha(\omega)} e^{- i \omega \tau} sin(\frac{\theta}{2}) e^{-i \phi_{r}}} {\hat{b}^{+}(\omega)}}{\hat{U}_{BS}}^{+} |vac\rangle_{5} | vac \rangle_{6} = \nonumber\\
e^{-\frac{1}{2} \int d\omega |\alpha(\omega)|^{2}} e^{{ \int d\omega {\alpha(\omega)} f(\omega){cos(\frac{\theta}{2})} e^{i \phi_{t}}(cos(\frac{\theta}{2})  e^{i \phi_{t}}  } {\hat{a}^{+}(\omega)} + sin(\frac{\theta}{2}) e^{-i \phi_{r}} {\hat{b}^{+}(\omega)}) } \nonumber \\
e^{{ \int d\omega {\alpha(\omega)} e^{- i \omega \tau} sin(\frac{\theta}{2}) e^{-i \phi_{r}}} ( -sin(\frac{\theta}{2}) e^{i \phi_{r}} {\hat{a}^{+}(\omega)}  + cos(\frac{\theta}{2}) e^{-i \phi_{t}} {\hat{b}^{+}(\omega)})}|vac\rangle_{2} |vac \rangle_{1}.
\end{gather}
     
\noindent In the next step we group according to the creation operators in appropriate modes:

\begin{gather}
     e^{-\frac{1}{2} \int d\omega |\alpha(\omega)|^{2}} 
     e^{{ \int d\omega {\hat{a}^{+}(\omega)} ({cos(\frac{\theta}{2})}^{2} e^{i 2 \phi_{t}}  } {\alpha(\omega)} f(\omega)  - 
      e^{- i \omega \tau} {{\alpha(\omega)}  sin(\frac{\theta}{2})}^{2} 
   ) } \nonumber\\
     e^{{ \int d\omega {\hat{b}^{+}(\omega)} } (    
     cos(\frac{\theta}{2}) sin(\frac{\theta}{2}) e^{i(\phi_{t}-\phi_{r})} {\alpha(\omega)} f(\omega)
     + {\alpha(\omega)} e^{- i \omega \tau} sin(\frac{\theta}{2}) cos(\frac{\theta}{2}) e^{-i(\phi_{t}+\phi_{r})}) }|vac\rangle_{2} |vac \rangle_{1} .
     \label{mat:S11}
\end{gather}{}

\noindent To obtain the state (\ref{mat:S11}) in a compact and useful form, we define two new variables $\gamma(\omega) = \alpha(\omega)  f(\omega)$ and $\zeta (\omega) = \alpha(\omega) e^{-i \omega \tau}$:

\begin{gather}
     e^{-\frac{1}{2} \int d\omega |\alpha(\omega)|^{2}} 
     e^{{ \int d\omega {\hat{a}^{+}(\omega)} ({cos(\frac{\theta}{2})}^{2} e^{i 2 \phi_{t}}  } \gamma(\omega)  - 
     \zeta (\omega)  {sin(\frac{\theta}{2})}^{2}
   ) } \nonumber\\
     e^{{ \int d\omega {\hat{b}^{+}(\omega)} } (    
     cos(\frac{\theta}{2}) sin(\frac{\theta}{2})  e^{i(\phi_{t}-\phi_{r})} \gamma(\omega)
     + \zeta (\omega) sin(\frac{\theta}{2}) cos(\frac{\theta}{2}) e^{-i(\phi_{t}+\phi_{r})}) }|vac\rangle_{2} |vac \rangle_{1} = \nonumber\\
      |\gamma cos^{2}(\frac{\theta}{2})e^{i 2 \phi_{t}}  -   \zeta sin^{2}(\frac{\theta}{2}) \rangle_{2} |\gamma cos(\frac{\theta}{2}) sin(\frac{\theta}{2}) e^{i(\phi_{t}-\phi_{r})} + \zeta cos(\frac{\theta}{2}) sin(\frac{\theta}{2}) e^{-i(\phi_{t}+\phi_{r})} \rangle_{1}.
      \label{mat:S13}
\end{gather}

\noindent The final form of the vector in formula \ref{mat:S13} after whole setup transformation can be written in presented form, if the sum of square module of the coefficients at the creation opertors is normalized to $|\alpha|^{2}$ (This is due to the definition of a coherent state formula \ref{mat:S4}). In the case of this system, this sum will be give $|\alpha|^{2}$ reduced by the losses related to the transmission of the object with the given function $f(\omega)$. So in order for formula \ref{mat:S13} to be true, the $|\alpha|^{2}$ coefficient should be scaled by losses and then the definition of the coherent state can be used.

\noindent  According to the picture (\ref{fig:setupz}c and \ref{fig:setupz}d) we will only take the state in port 2 from equation \ref{mat:S13}. That state propagated to the second beam-splitter (BS2), therefore the state before BS2 is :

\begin{gather}
 | \Omega \rangle_{2} |vac \rangle_{7} = \nonumber \\
     e^{-\frac{1}{2} \int d\omega |\Omega(\omega)|^{2}} 
     e^{{ \int d\omega {\hat{a}^{+}(\omega)} \Omega(\omega) } }  |vac\rangle_{2} |vac \rangle_{7},
     \label{mat:S14}
\end{gather}
where $\Omega = \gamma cos^{2}(\frac{\theta}{2})e^{i 2 \phi_{t}}  -   \zeta sin^{2}(\frac{\theta}{2}) $ . The state \ref{mat:S14} propagated to the beam-splitter 2, which splits this state between the two ports 5 and 6 (Fig. \ref{fig:setupz}d). Futhermore we assume that beam-splitters 1 and 2 have the same parameters ($\theta, \phi_{t}, \phi_{r}$). Similarly to the transformation by the first beam-splitter, we use the operator given by the formula \ref{mat:Sbs} and the properties of the unitary operator $\hat{U}_{BS} {\hat{U}_{BS}}^{+} = 1 $ : 
\begin{gather}
e^{-\frac{1}{2} \int d\omega |\Omega(\omega)|^{2}} 
     \Big( \hat{U}_{BS} e^{{ \int d\omega {\hat{a}^{+}(\omega)} \Omega(\omega) } } \hat{U}_{BS}^{+} \Big) \Big(\hat{U}_{BS}  |vac\rangle_{2} |vac \rangle_{7} \Big)
\end{gather}

\begin{gather}
e^{-\frac{1}{2} \int d\omega |\Omega(\omega)|^{2}} 
      e^{{ \int d\omega (cos(\frac{\theta}{2}) e^{i\phi_{t}}} {\hat{a}^{+}(\omega)} + sin(\frac{\theta}{2}) e^{-i \phi_{r}} {\hat{b}^{+}(\omega)} ) \Omega(\omega)  }   |vac\rangle_{5} |vac \rangle_{6} = \nonumber\\
       |\Omega cos(\frac{\theta}{2}) e^{i\phi_{t}} \rangle_{5} |\Omega sin(\frac{\theta}{2}) e^{-i \phi_{r}}  \rangle_{6}
\end{gather}

\noindent  Thus, the entire transformation performed by the system can be written as:

\begin{gather}
    \hat{U}_{ML}|\alpha \rangle_{1} |vac \rangle_{2} => \nonumber \\ |\gamma cos^{3}(\frac{\theta}{2})e^{i 3 \phi_{t}}  -   \zeta cos(\frac{\theta}{2}) sin^{2}(\frac{\theta}{2}) e^{i \phi_{t}} \rangle_{5} |\gamma cos(\frac{\theta}{2}) sin^{2}(\frac{\theta}{2}) e^{i(\phi_{t}-2\phi_{r})} + \zeta cos(\frac{\theta}{2}) sin^{2}(\frac{\theta}{2}) e^{-i(\phi_{t}+2\phi_{r})}  \rangle_{6} ,
\end{gather}

\noindent where $\hat{U}_{ML}$ is an operator that transforms through the entire system. For simplicity, we can write:

\begin{gather}
    \hat{U}_{ML} |\alpha\rangle_{1} |vac \rangle_{2} = > |\tilde{\alpha} \rangle_{5}
    |\tilde{\beta} \rangle_{6},
    \label{mat:S15}
\end{gather}

\noindent where $\tilde{\alpha} = \gamma cos^{3}(\frac{\theta}{2})e^{i 3 \phi_{t}}  -   \zeta cos(\frac{\theta}{2}) sin^{2}(\frac{\theta}{2}) e^{i \phi_{t}} $ and $ \tilde{\beta} =\gamma cos(\frac{\theta}{2}) sin^{2}(\frac{\theta}{2}) e^{i(\phi_{t}-2\phi_{r})} + \zeta cos(\frac{\theta}{2}) sin^{2}(\frac{\theta}{2}) e^{-i(\phi_{t}+2\phi_{r})} $.

\noindent We assume an ideal act of detection without losses and dark counts which is described by projector $ \hat{P}^{'} = \sum_{n=1}^{\infty}\frac{(\hat{a}^+)^{n}}{\sqrt{n!}}|vac \rangle\langle vac|\frac{(\hat{a})^{n}}{\sqrt{n!}} = 1 -|vac \rangle\langle vac|$. A detector described with such a measurement operator distinguishes the absence of a particle (this is described by the vacuum projection operator $|vac \rangle\langle vac|$) from any other particle number (this is described by $1$).  In the case of a detector described by operator $\hat{P}'$, the mean value of this operator in a coherent state is calculated as:

\begin{gather}
    \langle \alpha | \hat{P}^{'} | \alpha \rangle =  \langle \alpha | 1 - |vac \rangle \langle vac| | \alpha \rangle =1 - \langle \alpha | |vac\rangle \langle vac| | \alpha \rangle  = 1 - e^{- \int d\omega |\alpha(\omega)|^{2} }.
\end{gather}{}

\noindent However, the measurement performed in the experiment also measures the wavelength of light. For this reason, it must describe the vacuum state for a particular wavelength $ \hat{P}(\omega) = \sum_{n=1}^{\infty}\frac{(\hat{a}^+(\omega))^{n}}{\sqrt{n!}}|vac \rangle\langle vac| \frac{(\hat{a}(\omega))^{n}}{\sqrt{n!}}$. In the case of a detector described by operator $\hat{P}$, the mean value in a coherent state is calculated as:

\begin{gather}
     \langle \alpha | \hat{P}(\omega) | \alpha \rangle = 1 - e^{- |\alpha(\omega)|^{2} } .
\end{gather}

\noindent Using this knowledge, the probability of a coincident count detection $P_{cc}(\tau,\omega,\omega')$  can be calculated.A coincidence event is a simultaneous detection of particles by separate detectors.In this experiment, one detector is on path 5 and is described by operator:
\begin{gather}
    \hat{P}_{\tilde{\alpha}}(\omega) = \sum_{n=1}^{\infty}\frac{(\hat{a}^+(\omega))^{n}}{\sqrt{n!}}|vac \rangle_{5} {}_{5}\langle vac| \frac{(\hat{a}(\omega))^{n}}{\sqrt{n!}}.
\end{gather}
The second detector is located  on path 6 and is described by operator:
\begin{gather}
    \hat{P}_{\tilde{\beta}} (\omega') = \sum_{n=1}^{\infty}\frac{(\hat{b}^+(\omega'))^{n}}{\sqrt{n!}}|vac \rangle_{6} {}_{6}\langle vac| \frac{(\hat{b}(\omega'))^{n}}{\sqrt{n!}}.
\end{gather}
A coincidence event can be written using a tensor product of the two detection operator $\hat{P}_{\tilde{\alpha}}(\omega) \otimes \hat{P}_{\tilde{\beta}}(\omega')$.Thus, the mean value of the showned operator should be calculated in the state described by the formula \ref{mat:S15}:

\begin{gather}
   _{5} \langle \tilde{\alpha}| _{6} \langle
    \tilde{\beta}| \hat{P}_{\tilde{\alpha}} (\omega) \otimes \hat{P}_{\tilde{\beta}}(\omega') |\tilde{\alpha} \rangle_{5}
    | \tilde{\tilde{\beta}} \rangle_{6} = _{5}\langle \tilde{\alpha} | \hat{P}_{\tilde{\alpha}}(\omega) | \tilde{\alpha}\rangle_{5}  {}_{6} \langle \tilde{\beta} | \hat{P}_{\tilde{\beta}}(\omega') |\tilde{\beta} \rangle_{6} = \nonumber\\
   ( 1 - e^{- |\tilde{\alpha}(\omega) |^{2} })(1 - e^{-  |\tilde{\beta}(\omega')|^{2} }) = P_{cc}(\tau,\omega,\omega') .
\end{gather}{}

\end{document}